\documentclass{jetpl}
\twocolumn

\lat

\title{Current noise generated by spin imbalance in presence of spin relaxation}
\rtitle{}
\sodtitle{}
\author{V.S. Khrapai$^1$\,and\,K.E. Nagaev$^{1,2}$}
\rauthor{}
\sodauthor{}
\address{1. Institute of Solid State Physics RAS, ul Ak Osipyana 2, Chernogolovka 142432, Russia}
\address{2. Kotelnikov Institute of Radioengineering and Electronics, Mokhovaya 11-7, Moscow 125009, Russia}

\dates{\today}{*}
\abstract{We calculate current (shot) noise in a metallic diffusive conductor generated by spin imbalance in the absence of a net 
electric current. This situation is modeled in an idealized three-terminal setup with two biased ferromagnetic leads (F-leads) and one normal lead (N-lead). Parallel magnetization of the F-leads gives rise in spin-imbalance and finite shot noise at the N-lead. Finite spin relaxation results in an increase of the shot noise, which depends on the ratio of the length of the conductor ($L$) and the spin relaxation length ($l_s$). For $L\gg l_s$ the shot noise increases by a factor of two and coincides with the case of the anti-parallel magnetization of the F-leads.}

\begin{document}

\maketitle

\newcommand{\bn}{{\bf n}}
\newcommand{\bp}{{\bf p}}   
\newcommand{\br}{{\bf r}}
\newcommand{\bR}{{\bf R}}
\newcommand{\bk}{{\bf k}}
\newcommand{\bv}{{\bf v}}
\newcommand{\wk}{\omega_{\bf k}}
\newcommand{\nk}{n_{\bf k}}
\newcommand{\brho}{\boldsymbol{\rho}}
\newcommand{\eps}{\varepsilon}
\newcommand{\la}{\langle}
\newcommand{\ra}{\rangle}
\newcommand{\be}{\begin{eqnarray}}
\newcommand{\ee}{\end{eqnarray}}
\newcommand{\intl}{\int\limits_{-\infty}^{\infty}}
\newcommand{\dE}{\delta{\cal E}^{ext}}
\newcommand{\SE}{S_{\cal E}^{ext}}
\newcommand{\dsp}{\displaystyle}
\newcommand{\phit}{\varphi_{\tau}}
\newcommand{\p}{\varphi}
\newcommand{\tp}{\tilde p}
\newcommand{\uu}{\uparrow\uparrow}
\newcommand{\ud}{\uparrow\downarrow}
\newcommand{\du}{\downarrow\uparrow}
\newcommand{\dd}{\downarrow\downarrow}
\newcommand{\tG}{\tilde\Gamma}
\newcommand{\tP}{\tilde\Phi}
\newcommand{\fs}{f_{\sigma}}
\newcommand{\fms}{f_{-\sigma}}

The ability to detect nonequilibrium spin accumulation (imbalance) by all-electrical means is one of the key ingredients 
in spintronics~\cite{Zutic2004}. Transport detection typically relies on a nonlocal measurement 
of a contact potential difference induced by the spin imbalance by means of ferromagnetic contacts~\cite{Silsbee1980,Johnson1985,Jedema2001,Jedema2002,Lou2007} or spin resolving detectors~\cite{Ritchie2014}. A drawback of these approaches lies in a difficulty to extract the absolute value of the spin imbalance without an independent calibration.

An alternative concept of a spin-to-charge conversion via nonequilibrium shot noise was introduced in Ref.~\cite{Meair2011} and recently investigated experimentally~\cite{Kobayashi2015}. Here, the basic idea is that a nonequilibrium spin imbalance generates spontaneous current fluctuations, even in the absence of a net electric current. Being a primary approach~\cite{Blanter2000}, the shot noise based detection is potentially suitable for the absolute measurement of the spin imbalance. In addition, the noise measurement can be used 
for a local non-invasive sensing, as recently demonstrated with a semiconductor nanowire probe~\cite{SRep2016}.

It is well known, how a relaxation of the electronic energy distribution via inelastic electron-phonon~\cite{Nagaev1992} and electron-electron~\cite{Nagaev1995,Kozub1995} scattering influences the shot noise in diffusive conductors and nonequilibrium spin valves~\cite{Heikkila2013}. In this letter, we calculate the impact of a spin relaxation on the spin imbalance generated shot noise in the absence of inelastic processes. We find that the spin relaxation increases the noise up to a factor of two, depending on the ratio of the conductor length and the spin relaxation length.
%

\begin{figure}[t]
\center
\includegraphics[width=0.8\linewidth]{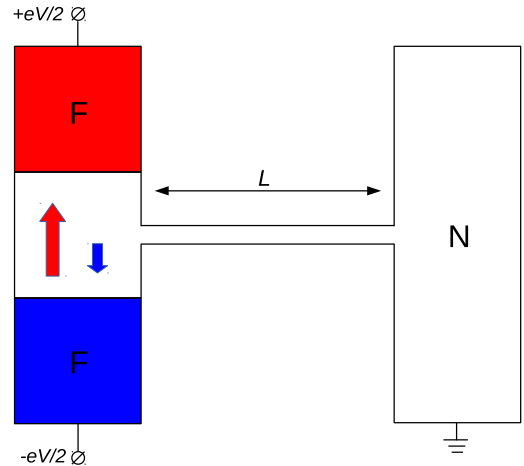}
\vspace{3mm}
\caption{Fig.~\ref{geometry}. The design of the system. A diffusive normal wire of the length $L$ is attached to 
normal islands on both ends. Nonequilibrium energy distribution on the left hand side of the wire generates the shot noise at a zero net current. The spin imbalance on the left-hand side of the wire is due to the electric current flowing from one ferromagnetic lead (red) to another one with opposite magnetization (blue).}
\label{geometry}
\end{figure}

\begin{figure}[h]
\center
\includegraphics[width=0.8\linewidth]{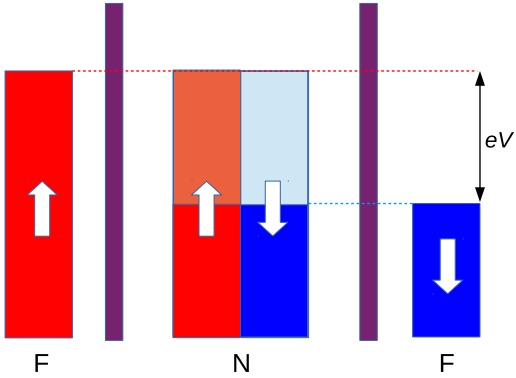}
\caption{Fig.~\ref{diagram}. Energy diagram of spin-up (red) and spin-down electrons (blue) in the ferromagnetic electrodes and in the normal-metal island at the left end of the wire. Stronger and lighter colours show  fully and partially occupied states, respectively. Narrower bars indicate tunnel barriers.}
\label{diagram}
\end{figure}

Consider the system shown in Fig.~\ref{geometry}. It consists of a diffusive wire, one end of which is grounded and the 
other is attached to a conducting island much larger than the transverse dimensions of the contact. The spin imbalance in the island is produced by electron tunneling through two junctions connecting it to ferromagnetic 
leads with antiparallel magnetizations. The junctions are assumed to have equal conductances much smaller than
that of the island, and the ferromagnetic leads are antisymmetrically biased by voltages $V/2$ and $-V/2$, so
the island has zero electrical potential and the net electrical current through the wire is zero. However
the tunneling results in a nonequilibrium spin-dependent distribution of electrons in the island (see 
Fig.~\ref{diagram}). If the conductance of the diffusive wire is small as compared with the conductances of tunnel junctions, the equation for the distribution functions of spin-up and spin-down electrons, $\fs$ $(\sigma = \pm)$ in the island may be written in the form
\begin{multline}
 \frac{\partial\fs}{\partial t}
 =
 \Gamma_{L\sigma}\,[f_0(\eps - eV/2) - \fs]
\\
 +
 \Gamma_{R\sigma}\,[f_0(\eps + eV/2) - \fs]
 -
 \frac{1}{\tau_{s}} (\fs - \fms),
 \label{f-eq}
\end{multline}
where $\Gamma$'s are the tunneling rates through the left and right
barriers for spin-up and spin-down electrons, $f_0(\eps)$ is the
equilibrium distribution function of electrons, and $\tau_{s}$ is the
spin-flip scattering time. 

The magnetization of the leads enters into Eq. (\ref{f-eq}) via the tunneling rates, which is nonzero for the
majority-spin electrons and zero for the minority-spin electrons.  For the antiparallel magnetization, we assume that $\Gamma_{L-} = \Gamma_{R+} = 0$ and $\Gamma_{L+} = \Gamma_{R-} = \Gamma$. the stationary solution of
Eq. (\ref{f-eq}) is given by
\begin{align}
 f_{+}(\eps) = \frac{1+\alpha}{2}\,f_0(\eps-eV/2) + \frac{1-\alpha}{2}\,f_0(\eps+eV/2),
 \label{f+}\\
 f_{-}(\eps) = \frac{1-\alpha}{2}\,f_0(\eps-eV/2) + \frac{1+\alpha}{2}\,f_0(\eps+eV/2),
 \label{f-}
\end{align}
where
\be
 \alpha = \frac{\Gamma\tau_s}{1 + \Gamma\tau_s}.
 \label{alpha}
\ee
For the parallel magnetization of the leads, $\Gamma_{L-} = \Gamma_{R-} = 0$ and $\Gamma_{L+} =
\Gamma_{R+} = \Gamma$, so $f_{+}(\eps)$ and $f_{-}(\eps)$ are given by Eqs. (\ref{f+}) and (\ref{f-})
with $\alpha=0$.

The distribution functions in the diffusive wire satisfy the diffusion equation
\be
 D\,\frac{\partial^2 f_{\sigma}}{\partial x^2} = \frac{f_{\sigma} - f_{-\sigma}}{2\,\tau_s}
 \label{diffusion-eq}
\ee
with the boundary condition at the left end given by Eqs. (\ref{f+}) and (\ref{f-}) and the 
boundary condition at the right end $f_{\sigma}(L,\eps) = f_0(\eps)$. The solution of this equation 
is of the form
\begin{multline}
 f_{\sigma}(x, \eps) = \frac{x}{L}\,f_0(\eps)
 \\
 +\frac{1}{2} \left(1 - \frac{x}{L}\right) [f_0(\eps-eV/2) + f_0(\eps+eV/2)] + 
 \\
 + \sigma\alpha\,\frac{ \sinh\!\left(\frac{L-x}{l_s}\right) }{2\,\sinh(L/l_s)}\,
 [f_0(\eps-eV/2) - f_0(\eps+eV/2)],
 \label{diffusion-sol}
\end{multline}
where $l_s = \sqrt{D\,\tau_s}$. 

The electrical noise is calculated as~\cite{Nagaev1992}
\be
 S_I = \frac{2}{R} \int_0^L \frac{dx}{L} \int d\eps\,
 \sum_{\sigma} f_{\sigma}\,(1 - f_{\sigma}).
 \label{S_I-1}
\ee
A substitution of Eq. (\ref{diffusion-sol}) into Eq. (\ref{S_I-1}) gives at low
temperatures $T \ll eV$
\begin{multline}
 S_I = \frac{eV}{R}\,\Biggl[ \frac{2}{3} 
     - \frac{\alpha^2 l_s}{2 L}\,\frac{e^{2L/l_s} + 1}{e^{2L/l_s} - 1}
     + \frac{2\alpha^2 e^{2L/l_s} }{(e^{2L/l_s} - 1)^2}
     \Biggr].
 \label{S_I-2}
\end{multline}

\begin{figure}[t]
\includegraphics[width=0.9\linewidth,bb=0 0 550 550]{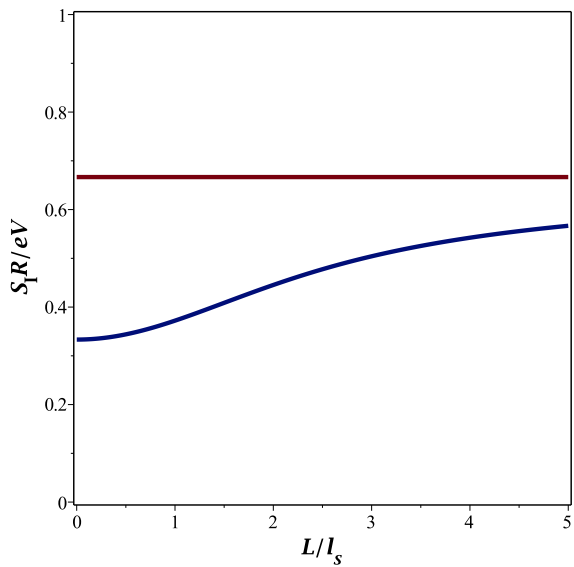}
\caption{Fig.~\ref{S-L}. The dependence of the dimensionless spectral density $S_I R/eV$ on the ratio
$L/l_s$ for $\alpha=0$ (red curve) and $\alpha=1$ (blue curve).}
\label{S-L}
\end{figure}

In Fig.~\ref{S-L} we plot the dependence of dimensionless shot noise spectral density as a function of the conductor length (in units of $l_s$). The upper curve corresponds to the case of $\alpha=0$, i.e. the non-equilibrium double-step electronic energy distribution in the absence of spin imbalance, see eqs.~(\ref{f+}) and (\ref{f-}). As expected, the spin relaxation has no effect here and we recover a familiar result~\cite{SRep2016,Sukhorukov1999} $S_I= 2eV/3R$. By contrast, the case of $\alpha=1$, which corresponds to the injection of a pure spin current into the conductor, is strongly sensitive to the spin relaxation. The lower curve in Fig.~\ref{S-L} demonstrates that in this case the noise equals $S_I=eV/3R$ for a vanishing spin relaxation and increases with the ratio $L/l_s$. In the asymptotic limit $L\gg l_s$, the noise increases by a factor of 2, see eq.~(\ref{S_I-2}), where the results for $\alpha=0$ and $\alpha=1$ coincide. 

Our results lead to a more general conclusion. Elastic spin relaxation always tends to equalize the distributions of spin-up and spin-down electrons and bring them to $\overline{f}_{\pm}=(f_++f_-)/2$. Obviously, this results in the increase of the shot noise, since $2\overline{f}_{\pm}(1-\overline{f}_{\pm})\geq \sum f_\sigma(1-f_\sigma)$, cf. eq.~(\ref{S_I-1}). The magnitude of the increase is thus a measure of the spin imbalance.

In summary, we calculated the impact of the spin relaxation on the nonequilibrium shot noise generated by spin imbalance in a diffusive conductor. In an idealized three-terminal setup with two tunnel-coupled ferromagnetic leads and one normal lead the noise is found to increase by a factor of two as the length of the conductor becomes larger than the spin relaxation length. The increase of the noise governed by the spin relaxation is a generic effect that may be useful for the measurements of the spin relaxation length and the degree of the spin imbalance. 

We acknowledge discussions with V.T. Dolgopolov, S.V. Piatrusha and E.S. Tikhonov.

\end{document}